\documentclass[aoas,nameyear,seceqn,dvips]{arximspdf}
\usepackage{graphics}

\doi{10.1214/08-AOAS221}
\volume{4}
\issue{1}
\pubyear{2010}
\firstpage{5}
\lastpage{25}

\makeatletter
\DeclareMathAlphabet\mathcaligr{OMS}{cmsy}{m}{n}
\renewcommand{\mathcal}{\mathcaligr}
\newcommand{\cal}{\mathcaligr}
\renewcommand{\cite}{\citet}
\newcommand{\implies}{\Longrightarrow}

\newcommand{\IE}{\mathbb{E}}
\newcommand{\IVar}{\mathbb{V}}

\newproclaim{Definition}{Definition}
\newproclaim{Result}{Result}

\newcommand{\vZ}{{g}}
\newcommand{\veta}{\eta}
\newcommand{\vxi}{\xi}
\newcommand{\itall}{Y}
\makeatother

\begin{document}
\begin{frontmatter}

\title{Modeling social networks from sampled data\protect\thanksref{T1}}
\runtitle{Modeling social networks from sampled data}
\pdftitle{Modeling social networks from sampled data}

\begin{aug}
\author[A]{\fnms{Mark S.} \snm{Handcock}\corref{}\ead[label=e1]{handcock@stat.washington.edu}}
\and
\author[B]{\fnms{Krista J.} \snm{Gile}\ead[label=e2]{krista.gile@nuffield.ox.ac.uk}}
\runauthor{M. S. Handcock and K. J. Gile}
\affiliation{University of California--Los Angeles and Nuffield College}
\address[A]{Department of Statistics\\
University of California\\
Los Angeles, California 90095-1554\\
USA\\
\printead{e1}} 
\address[B]{Nuffield College\\
University of Oxford\\
New Road\\
Oxford 0X1 1NF\\
United Kingdom\\
\printead{e2}}
\end{aug}

\thankstext{T1}{Supported by NIH
Grants R01 DA012831 and R01 HD041877, NSF Grant
MMS-0851555 and ONR Grant N00014-08-1-1015.}

\received{\smonth{12} \syear{2007}}
\revised{\smonth{9} \syear{2008}}

%
\begin{abstract}
Network models are widely used to represent relational
information among interacting units and the structural implications of
these relations. Recently, social network studies have focused a great
deal of attention on
random graph models of networks whose nodes
represent individual social actors and whose edges represent a
specified relationship between the actors.

Most inference for social network models assumes that the presence or
absence of all possible links is observed, that
the information is completely reliable, and that there are no measurement
(e.g., recording) errors. This is clearly not true in practice, as
much network data is collected though sample surveys. In
addition even if a census of a population is attempted, individuals
and links between individuals are missed (i.e., do not appear in the
recorded data).

In this paper we develop the conceptual and computational theory for
inference based on sampled network information. We first review forms
of network sampling designs used in practice.
We consider inference from the
likelihood framework, and develop a typology of network data that
reflects their treatment within this frame. We then develop inference
for social network models based on information from adaptive network
designs.

We motivate and illustrate these ideas by analyzing the effect of link-tracing
sampling designs on a collaboration network.
\end{abstract}

%

\end{frontmatter}

\section{Introduction}\label{intro}

%
%
%
%
Networks are a useful device to represent ``relational data,''
that is, data with properties beyond the attributes of the
individuals (nodes) involved. Relational
data arise in many fields and network models
are a natural approach to representing the 
patterns
of the relations between nodes.
Networks can be used to describe such diverse ideas as the
behavior of epidemics, the interconnectedness of 
corporate boards, and
networks of genetic regulatory interactions.
In social network applications, the nodes in a graph
typically
represent
individuals, and the ties (edges)
represent a specified relationship between individuals. Nodes
can also be used to represent larger social units (groups,
families, organizations), objects (airports, servers,
locations) or abstract entities (concepts, texts, tasks,
random variables).
We consider here stochastic models for such graphs. These
models attempt to represent the stochastic mechanisms that
produce relational ties, and the complex dependencies thus induced.

Social network data typically consist of a set of $n$ actors
and a relational tie random variable, $Y_{ij}$, measured on each possible
ordered pair of actors, $(i,j)$, $i,j=1,\ldots,n, i\neq j$.
In the most simple cases, $Y_{ij}$ is
a dichotomous variable, indicating the presence or absence of some
relation of interest, such as friendship, collaboration, transmission
of information or disease, etc. The data are often represented by
an $n\times n$ sociomatrix $Y$, with diagonal elements, representing
self-ties, treated as structural zeros. In the case of binary relations,
the data can also be thought of as a graph in which the nodes are
actors and the edge set is $\{ (i,j)  \dvtx Y_{ij}=1 \}$.
For many networks the relations are undirected in the sense that
$Y_{ij}=Y_{ji}, i,j=1,\ldots,n.$

In the application in this
paper we consider a network formed from the
collaborative working relations between $n=36$ partners in a New
England law
firm [\citet{laz01}].
We focus on the undirected relation where a tie is said to
exist between two partners if and only if both indicate that
they collaborate with the other.
The scientific objective is to explain the observed structural pattern
of collaborative ties as a function of nodal and relational
attributes.
The relational data is supplemented
by four actor attributes: seniority (the rank number of chronological
entry into the firm divided by 36), practice
(there are two possible values, litigation${}={}$0 and corporate law${}={}$1),
gender (3 of the 36 lawyers are female) and office (there are three
different offices in three different cities each of different size).


For large or hard-to-find populations of actors it is difficult to obtain
information on all actors and all relational ties. As a result, various
survey sampling strategies and methods are applied. Some of these
methods make use of
network information revealed by earlier stages of sampling to guide
later sampling. These adaptive designs
allow for more efficient sampling than conventional sampling designs. We
consider such designs in Section~\ref{mbs}.

Most of the work presented here considers the network over the set of
actors to
be the realization of a stochastic process. We seek to model that
process. An
alternative is to view the network as a fixed structure about which
we wish to make inference based on partial observation.

In this paper we develop a theoretical framework for inference from
network data that are partially-observed due to sampling.
This work extends the
fundamental work of \citet{thompsonfrank2000}.
For purposes of presentation, we focus on the
relational data itself and suppress reference to\break covariates
of the nodes. This more general situation is dealt with in\break
\citet{HanGile07}.

In Section~\ref{mbs} we present a conceptual framework for network
sampling.
We extend this framework in Section~\ref{inference} to focus on inference from
sampled network data.
We first consider the limitations of design-based inference in this
setting, then focus on likelihood-based inference. 
Section~\ref{sec4} presents the rich Exponential Family Random
Graph Model (ERGM) family of models that has been applied to complete network
data.
Section~\ref{sec5} presents a study of the effect of sampling from
a known complete network of law firm
collaborations.
Finally, in Section~\ref{sec6}, we discuss the overall ramifications for the
modeling of
social networks with sampled data and note some extensions.


\section{Network sampling design}\label{mbs}

In this section we consider the conceptual and computational
theory of network sampling.

There is a substantial literature on network sampling designs.
Our development here follows \cite{thompsonseber1996} and
\cite{thompsonfrank2000}.
Let~${\cal Y}$ denote the set of possible networks on the
$n$ actors.
Note that in most network samples, the unit of sampling is the actor or
node, while the unit of analysis is typically the dyad.
Let $D$ be the $n\times n $ random binary matrix indicating if the
corresponding element of $Y$ was sampled or not. The value of the
${i, j}$th element is 0 if the $(i, j)$ ordered pair was
not sampled and $1$ if the element was sampled.
Denote the sample space of $D$ by $\cal{D}.$
We shall refer to the probability
distribution of $D$ as the \textit{sampling design.} The sampling
design is often related to the structure of the graph
and a parameter $\psi\in\Psi,$
so we posit a model for it. Specifically,
let $P(D=d |Y=y;\psi)$ denote the probability of
selecting sample $d$ given a network~$y$ and parameter $\psi.$ 

Under many sampling designs the set of sampled dyads
is determined by the set of sampled nodes. Let $S$
represent a binary random $n$-vector indicating a subset of the nodes,
where the
$i$th element is $1$ if the $i$th node is part of the set,
and is $0$ otherwise. We 
often
consider situations where $D$ is
determined by some $S$ which is itself a result of a sample design
denoted by $P(S \vert Y, \psi).$
For example, consider an
undirected
network where
the set of observed dyads are those that are incident on at least
one of the sampled nodes. In this case
$D={S\circ 1} + {1\circ S} - {S\circ S},$
where~$1$ is the binary $n$-vector of $1$s.
A primary example of this is where people are sampled and surveyed
to determine all their edges.

We introduce further notation to allow us to refer to the
observed and unobserved portions of the relational 
structures. Denote the observed
part of the complete graph $Y$ by
$Y_{\mathrm{obs}}= \{{Y_{ij}  \dvtx D_{ij} =1} \} $ and the unobserved part
by $  Y_{\mathrm{mis}}  = \{
{Y_{ij}  \dvtx D_{ij} =0}  \}. $
The full \textit{observed data} is then
$ \{ {Y_{\mathrm{obs}}, D}  \},$
in contrast to the
\textit{complete data}: $ \{ {Y_{\mathrm{obs}}, Y_{\mathrm{mis}}, D} \}.$ We
will write the complete
graph $Y =  \{ {Y_{\mathrm{obs}}  ,Y_{\mathrm{mis}}  }  \}.$
%
In addition, we make the convention that undefined numbers act as
identity elements in addition and multiplication. So a number $x$ plus
or multiplied by
an undefined number $y$
is $x$, and hence $Y=Y_{\mathrm{obs}}+Y_{\mathrm{mis}}.$
For a given network $y\in{\cal Y},$
denote the corresponding data
as $ \{ {y_{\mathrm{obs}}, d}  \}$ and the other elements by their
lower-case versions $y=y_{\mathrm{obs}}+y_{\mathrm{mis}}.$
Finally denote ${\cal Y}(y_{\mathrm{obs}})=\{v  \dvtx y_{\mathrm{obs}}+v \in{\cal Y} \},$
that is the set of possible unobserved elements which together with
$y_{\mathrm{obs}}$ result in valid network. The set $y_{\mathrm{obs}}+{\cal Y}(y_{\mathrm{obs}})$
is then the restriction of ${\cal Y}$ to $y_{\mathrm{obs}}$.

A sampling design is \textit{conventional} if it does not use
information collected during the survey to direct subsequent sampling
of individuals (e.g., network census and ego-centric designs).
Specifically, a design is conventional if
$P(D=d | Y=y;\psi)=P(D=d | \psi)\ \forall y\in{\cal Y}.$
A simple example of a conventional sampling design for networks
is an \textit{ego-centric design}, consisting of a simple random sampling
of a subset of the actors, followed by
complete observation of the dyads originating from those actors.
A complete census of the network is another.
More complex examples include designs using probability sampling of pairs
and auxiliary variables.
Alternatively, we call a sampling design \textit{adaptive} if it uses
information collected during the survey to direct subsequent
sampling, but the sampling design depends only on the observed data.
Specifically, a design is adaptive if:
$P(D=d |Y=y;\psi)=P(D=d| Y_{\mathrm{obs}}=y_{\mathrm{obs}}, \psi)\ \forall y\in
y_{\mathrm{obs}}+{\cal Y}(y_{\mathrm{obs}}).$
Hence a design can be adaptive for a given $y_{\mathrm{obs}}$ (rather than all possible
observed data), although most common such designs are adaptive for all possible
data observed under them.
Conventional designs can be considered to
be special cases of adaptive designs.

Note that adaptive sampling designs satisfy
%
\begin{equation}
P(D=d \vert Y_{\mathrm{obs}} ,Y_{\mathrm{mis}}, \psi)=P(D=d \vert Y_{\mathrm{obs}}, \psi),
\label{eqn:mar}
\end{equation}
a condition called ``missing at random'' by
\citet{Rub76} in the context of missing data.
Note that this is a bit misleading---it does not say that the
propensity to be observed is unrelated
to the unobserved portions of the network, but that this relationship
can be
explained by the data that are observed. The observed part of the data are
often
vital to equality~(\ref{eqn:mar}). Hence adaptive designs are
essentially those
for which the unobserved dyads are missing at random.




Denote by $ [ a  ]$ the vector-valued
function that is $1$ if the corresponding element of the vector $ a $
is logically true, and
$0$ otherwise. Let $a\times{b}$ be the elementwise product of the
column vector $a$ and
the column vector $b$ and $a\cdot{b}$ be the scalar product
$\sum_{j}{a_{j}b_{j}}$.
Let $a\circ{b}$ be the outer product matrix with $ij$th element
$a_{i}b_{j}$.
If $y$ is a matrix and $b$ a vector let
$y\cdot{b}$ be the column vector with $i$th element
$\sum_{j}{y_{ji}b_{j}}$.

\subsection{Some adaptive designs for undirected networks}

We now consider several examples of adaptive designs for undirected networks.

\subsubsection{Example: Ego-centric design}\label{sec2.1.1}
Consider a simple \textit{ego-centric design}:
\begin{enumerate}
\item Select individuals at random, each with probability $\psi.$
\item Observe all dyads involving the selected individuals (i.e.,
dyads with at least one of the selected individuals as one of the pair of
actors).
\end{enumerate}

The sampling design can be determined for this case. First note that
%
\[
 P(D_{ij} =1 \vert Y, \psi)=1-(1-\psi)^2 \qquad \forall i\ne j.
\]
This, however, does not give the joint distribution of $D.$
Let $S$ be the binary $n$-vector where 1 and 0 indicate that
the corresponding individual has been selected, or not, respectively. Within\vspace*{1pt}
this design, $ S $ is determined by $D$ (i.e., $ S = [
{D1=(n-1)1}  ]) $. Then $ P(S =s \vert Y,\psi) =\psi^{1\cdot
s}(1-\psi
)^{n-1\cdot s}$,  $s\in \{ {0,1}  \}^n $. If the \textit
{i}th element of $ S $ is 1
then all elements in the \textit{i}th row and column of $D$ are 1.
$D_{ij} = 0$ if and only if both the \textit{i}th and \textit{j}th
elements of $S$ are both 0.
Hence the probability distribution of $D$ is
\[
 P(D=d \vert Y, \psi)=\psi^{1 \cdot s}(1-\psi
)^{n-1 \cdot s}
\]
for
\[
 d={1\circ s}+{s\circ 1}-{s\circ s},\qquad  s\in \{ {0,1}
 \}^n .
\]
Note that the distribution does not depend on $Y$, and is therefore
conventional.

\subsubsection{Example: One-wave link-tracing design}\label{sec2.1.2}
We refer to any sample in which subsequent nodes are enrolled based on
their observed relations with other sampled nodes as a \textit
{link-tracing} design. Consider the one-wave link-tracing design
specified as follows:
\begin{enumerate}
\item Select individuals at random, each with probability $\psi.$
\item Observe all dyads involving the selected individuals.
\item Identify all individuals reported to have at least one relation
with the initial sample, and select them with probability $1$.
\item Observe all dyads involving the newly selected individuals.
\end{enumerate}

Let $S_0$ denote the indicator vector for the initial sample and $S_1$
the indicator for the added individuals not in the initial sample. Then the
whole sample of individuals is $S=S_0+S_1$. As in the undirected
ego-centric design, $D={1\circ S}+{S\circ 1}-{S\circ S}$. Note
that $S_1 = [ {YS_0\times(1-S_0)} > 0 ] $ is derivable from
$S_0$ and $Y$. Hence
%
%
\[
P(D=d \vert Y, \psi)=\sum_{s_0 :{\kern1pt}{\kern1pt}s_0
+ [ {Ys_0\times(1-s_0)} > 0 ]=s} {\psi^{1 \cdot s_0
}(1-\psi)^{n-1 \cdot s_0 }}
\]
for
\[
 d={1\circ s}+{s\circ 1}-{s\circ s},\qquad  s\in \{ {0,1}
 \}^n .
\]

\subsubsection{Example: Multi-wave link-tracing design}
Consider a \textit{multi-wave link-tracing design}
in which the complete set of partners of the $k$th wave
are enrolled, that is, the link-tracing process described above is
carried out $k$ times. If $k$ is fixed in advance this is called \textit{$k$-wave link-tracing}.

Let $S_0 $ denote the indicator for the initial sample, $ S_1
$ the indicator for the added individuals in the first wave not in the
initial sample,\,\ldots, $ S_k $ the indicator for the added
individuals in
wave $ k $ not in the prior samples. Then the whole sample of
individuals is
$S=S_0 +S_1 +\cdots+S_k.$ As in the ego-centric design
$D={1\circ S}+{S\circ 1}-{S\circ S}$. Note that
$S_m = [{YS_{m-1}\times(1-\sum_{t=0}^{m-1}S_{t})}>0
]$, $m=1,\ldots,k $ is derivable from $ S_0 $ and
$ Y.$
Then
%
\[
P(D=d \vert Y,\psi)=\sum_{s_0 :{\kern1pt} s_0 +s_1
+\cdots+s_k =s} {\psi^{1 \cdot s_0 }(1-\psi)^{n-1\cdot s_0 }}
\]
for
$
d=1\circ s+s\circ 1-s\circ s$,  $s\in \{ {0,1}
\}^{n}.
$
Here $S_m = [
YS_{m-1}\times(1-\sum_{t=0}^{m-1}S_{t})>\break0 ]= [
Y_{\mathrm{obs}} S_{m-1}\times(1-\sum_{t=0}^{m-1}S_{t}) > 0
 ]$,  $m=1,\ldots,k $ so that the individuals selected in the
successive waves only depend on the observed part of
the graph, and not on the unobserved portions of the graph. Clearly,
this is also true for one-wave link-tracing as a simple case of
$k$-wave link-tracing.
Note
that it may be possible that $S_m =\varnothing$ for some $m<k,$ so
that subsequent
waves do not increase the sample size (i.e., $S_{k} = \varnothing).$ A
variant of
the $k$-wave link-tracing design is the \textit{saturated link-tracing}
design, in which sampling
continues until wave $m$, such that $S_m = \varnothing$. We interpret
$k$ as the bound on the number of waves sampled imposed by the sampling
design. Since saturated link-tracing does not restrict the number of
waves sampled, we represent it by setting $k=\infty.$

\subsection{Some adaptive designs for directed networks}
We can also consider variants of these adaptive designs for directed networks.

\subsubsection{Example: Ego-centric design}
Consider a simple \textit{ego-centric design}:
\begin{enumerate}
\item Select individuals at random, each with probability $\psi.$
\item Observe all directed dyads originating at the selected individuals.
\end{enumerate}

As before, the sampling design can be determined for this case.
Since a directed dyad is observed only if its tail node is sampled,
\[
 P(D_{ij} =1 \vert Y, \psi)=\psi\qquad  \forall i\ne j
\]
and
$D={S_{0}\circ 1}$.
Hence the probability distribution of $D$ is
\[
 P(D=d \vert Y, \psi)=\psi^{1\cdot s}(1-\psi
)^{n-1\cdot s}
\]
for
$
d={s\circ 1},  s\in \{ {0,1}  \}^n
$
and the distribution does not depend on $Y$. As in the undirected case,
this design is therefore conventional.

\subsubsection{Example: One-wave link-tracing design}
Consider a one-wave link-tracing design on a directed network
specified as follows:
\begin{enumerate}
\item Select individuals at random, each with probability $\psi.$
\item Observe all directed dyads originating at the selected individuals.
\item Identify all individuals receiving an arc from a member of the
initial sample, and select them with probability $1$.
\item Observe all directed dyads originating at the newly selected individuals.
\end{enumerate}

Let $S_0$ denote the indicator vector for the initial sample and $S_1$
the indicator for the added individuals not in the initial sample. Then the
whole sample of individuals is $S=S_0+S_1$. As in the ego-centric design
$D={S\circ 1}$ and
\[
P(D=d \vert Y, \psi)=\sum_{s_0 :{\kern1pt}{\kern1pt}s_0
+ [ {Y s_0 \times(1-s_0) >0}  ]=s} {\psi^{1\cdot s_0
}(1-\psi)^{n-1\cdot s_0 }}
\]
for
$
d={s\circ 1},  s\in \{ {0,1}  \}^{n}.
$

\subsubsection{Example: Multi-wave link-tracing design}
Consider a directed version of the
multi-wave link-tracing design in which
the complete set of out-partners of the \textit{k}th wave
are enrolled.
The whole sample of individuals is
$S=S_0 +S_1 +\cdots+S_k.$
And
$ S_m = [ {{Y\cdot S_{m-1}}\times(1-\sum_{t=0}^{m-1}S_{t}) }
>0  ]$, $ m=1,\ldots,k $ is derivable from $ S_0 $ and
$ Y.$
Then
%
\[
P(D=d \vert Y,\psi)=\sum_{s_0 :{\kern1pt} s_0 +s_1
+\cdots+s_k =s} {\psi^{1\cdot s_0 }(1-\psi)^{n-1\cdot s_0 }}
\]
for
$d={s\circ 1},  s\in \{ {0,1}  \}^n,$ where we
note that
$S_m = [ {Y\cdot S_{m-1}}\times(1-\sum_{t=0}^{m-1}S_{t})>0
 ] = [ {{Y_{\mathrm{obs}}\cdot S_{m-1}}\times(1-\sum
_{t=0}^{m-1}S_{t})>0} ],$
$m=1,\ldots,k$ so that the individuals selected in
successive waves of depend only on
the previously observed part of the graph, and not on the unobserved portions.
The saturated link-tracing design is represented by $k=\infty.$

\section{Inferential frameworks}\label{inference}

In this section we consider two frameworks for inference based on
sampled data.
In the \textit{design-based} framework $y$ represents the fixed population
and interest focuses on characterizing $y$ based on partial observation.
The random variation considered is due to the sampling design alone. A~key advantage
of this approach is that it does not require a model for
the data
themselves, although a model may also be used to guide design-based inference
[\cite{sarndal1992}].
Under the 
\textit{model-based} 
framework, $Y$~is
stochastic and is a realization from a stochastic process
depending on a parameter $\veta.$
Here interest focuses on $\veta$ which characterizes the
mechanism that produced the complete network $Y$.
We find severe limitations of the design-based framework for data from
link-tracing samples, and focus on likelihood inference within the model-based
framework. 

\subsection{Design-based inference for the network}\label{designinference}
In the design-based frame, the unobserved data values, or some
functions thereof, are analogous to the parameters of interest in
likelihood inference. The population of data values is treated
as fixed, and all uncertainty in the estimates is due to the sampling
design, which is typically assumed to be fully known (not just
up to the parameter $\psi$).

Inference typically focuses on identifying design-unbiased estimators
for quantities of interest measured on the complete network.
In an undirected network analysis setting, for example, we
can consider\vspace*{1pt} estimating $\tau= \sum_{i < j} y_{ij}$, the number of
edges in the network. Note that $y$ is a partially-observed matrix of
constants in this setting. Then $\hat{\tau}$ is design-unbiased for
$\tau$ if
\[
\IE_{D}[\hat{\tau}\vert\psi, y] = \tau,
\]
where the expectation is taken over realizations of the sampling process.
Specifically,
\[
\IE_{D}[\hat{\tau}(Y_{\mathrm{obs}},D)\vert\psi,y] = \sum_{d \in\cal{D}}
\hat{\tau}(y_{\mathrm{obs}}(d),d) P(D=d \vert\psi, y),
\]
where $\hat{\tau}(y_{\mathrm{obs}}(d),d)$ is the estimator expressed as a
function of
the observed network information.
Similarly, the variance of the estimator is computed with respect to
the variation induced by the sampling procedure
\[
\IVar_{D}[\hat{\tau}(Y_{\mathrm{obs}},D)\vert\psi,y] = \sum_{d \in\cal
{D}} \bigl(\hat{\tau}(y_{\mathrm{obs}}(d),d)-\tau\bigr)^2 P(D=d \vert\psi,y).
\]

The Horvitz--Thompson estimator is a classic tool of design-based
inference, and is based on inverse-probability weighting the sample.
In our example, it is
\[
\hat{\tau}(Y_{\mathrm{obs}},D)=\sum_{i<j : D_{ij}=1} \frac{y_{ij}}{\pi_{ij}},
\]
where the
\textit{dyadic sampling probability} $\pi_{ij}=P(D_{ij}=1 \vert\psi,
y)$ is the probability of observing dyad $(i,j)$.

Consider an estimator of $\tau$ based on relations observed
through the ego-centric design of Section~\ref{sec2.1.1}.
%
Then
\[
\pi_{ij} = 1- (1-\psi)^{2} \qquad  \forall   i,j.
\]
The classic Horvitz--Thompson estimator $\hat{\tau}$ of $\tau$ then
weights each observation by the inverse of its sampling probability
\[
\hat{\tau}=\sum_{i<j : D_{ij}=1} \frac{y_{ij}}{\pi_{ij}}    =
\frac{1}{1-(1-\psi)^2} \sum_{i<j : D_{ij}=1} y_{ij} .
\]
Then
\[
\IVar(\hat{\tau}) = \sum_{i<j} \sum_{k<l}  \{
[1-(1-\psi)^2]^{-2}\pi_{ij,kl} -1
 \}
y_{ij} y_{kl},
\]
%
where $\pi_{ij,kl} = P(S_{0i}+S_{0j} >0, S_{0k}+S_{0l} >0 )$ or
%
\[
\pi_{ij,kl} =
\cases{ \pi_{ij}, &\quad $i=k$,  $j=l$,\cr
\pi_{ij} \pi_{kl}, &\quad $i \notin\{k,l\}$   and   $j \notin\{k,l\}$,
\vspace*{2pt}\cr
\psi^3 -3\psi^2, &\quad  otherwise. 
}
\]
Among the many available estimators for the variance of the
Horvitz--Thompson estimator is the Horvitz--Thompson variance estimator:
\[
\hat{\IVar}(\hat{\tau}) = \sum_{i<j: D_{ij}=1} \sum_{k<l:
D_{kl}=1} \frac{1}{\pi_{ij,kl}} \{
[1-(1-\psi)^2]^{-2}\pi_{ij,kl} -1
 \}
y_{ij} y_{kl}.
\]

Note the importance of the unit sampling probabilities in these
estimators. This is a hallmark of design-based inference:
inference relies on full knowledge of the sampling procedure in
order to make unbiased inference without making assumptions about
the distribution of the unobserved data. This typically requires
knowledge of the sampling probability of each unit in the sample.
This procedure is complicated in the network context, in
that we require the sampling probabilities of the units of
analysis, dyads, which are different from the units of sampling,
nodes. In fact, for even single-wave link-tracing samples, the
dyadic sampling probabilities are not observable.

To see this, define the \textit{nodal neighborhood of a dyad} $(i,j)$,
$N(i,j)$, where
$k \in N(i,j) \iff\{ S_{0k} =1 \implies D_{ij} = 1$\}. Then $\pi_{ij} =
P(\exists k    \dvtx  S_{0k} = 1, k \in N(i,j))$.


For the one-wave link-tracing design of Section~\ref{sec2.1.2},
$N(i,j) = \{k\} \dvtx y_{ik}=1 \textrm{ or } y_{jk}=1 \textrm{ or }
k \in\{i,j\} $. Then if the initial sample $S_0$ is drawn according
to the design in Section~\ref{sec2.1.2},
$\pi_{ij} = 1- (1-\psi)^{\Vert N(i,j)\Vert}$.
Suppose $S_{0i}=1$, and $S_{0j}=0$. Then dyad $(i,j)$ is observed, but
$\Vert N(i,j)\Vert$ is unknown because it is unknown which $k$ satisfy
$y_{jk} = 1$. The link-tracing sampling structures for which
nodal and dyadic sampling probabilities are observable are
summarized in Table~\ref{table:obsprobs}. For directed networks,
we assume sampled nodes provide information on their out-arcs
only, so that $D$ is not symmetric and $D_{ij} = 1
\iff S_i=1$.



\begin{table}
\caption{Observable sampling probabilities under various sampling
schemes for directed and undirected networks.
Nodal and dyadic sampling probabilities are considered separately. ``X''
indicates observable sampling probabilities, while a blank indicates
unobservable sampling probabilities}
\label{table:obsprobs}
\begin{tabular*}{\textwidth}{@{\extracolsep{\fill}}lcccc@{}}
\hline
 & \multicolumn{2}{c}{\textbf{Nodal probabilities} $\bolds{\pi_i}$} &
\multicolumn{2}{c@{}}{\textbf{Dyadic probabilities} $\bolds{\pi_{i\!\! j}}$}\\[-5pt]
\textbf{Sampling} & \multicolumn{2}{c}{\hrulefill} &
\multicolumn{2}{c@{}}{\hrulefill}\\
\textbf{scheme} & \textbf{Undirected} & \textbf{Directed} & \textbf{Undirected} & \textbf{Directed}\\
\hline
Ego-centric & X & X & X & X\\
One-wave & X   \\
$k$-wave, $1<k<\infty$  \\
saturated & X  \\
\hline
\end{tabular*}
\end{table}

Of the designs considered here, dyadic sampling probabilities are observable
only for ego-centric samples, and 
never for link-tracing designs. Nodal sampling probabilities are also
observable for ego-centric sampling, as well as for one-wave and saturated
link-tracing designs in undirected networks.
Overall, this table presents strong limitations to the
applicability of design-based methods requiring the knowledge of
sampling probabilities to link-tracing designs. Note that this
limitation is not specific to dyad-based network statistics.
Estimation of triad-based network statistics such as a triad
census would be subject to similar limitations. A
Horvitz--Thompson style estimator would rely on a weighted sum of
observed triads, weighted according to sampling probabilities.
Sampling probabilities for triads would be even more complex, as
they would typically require sampling of two of the three nodes
involved in an undirected case, and at least two of the three
nodes in an directed case, depending on the triad census. Both
of these sampling probabilities would not be possible to compute
for link-tracing samples in which the degrees or in-degrees of
some involved nodes are unobserved.

Not surprisingly, most of the work on design-based estimators for
link-tracing samples has focused on the cases where sampling
probabilities are observable: typically for one-wave or saturated
samples used to estimate population means of nodal covariates.
\cite{frank2005} presents a good overview and
extensive citations to this literature. See also \cite{thompsoncollins2002}; \cite{snijders1992}. Although examples
tend to focus on instances where sampling probabilities are
observable, the limited applicability of classical design-based
methods in estimating structural network features based on
link-tracing samples has not been emphasized in the literature.




In the absence of observable sampling probabilities, design-based inference
requires a mechanism for estimating sampling probabilities. This is
most often
necessary in the context of out-of-design missing data, and addressed with
approaches such as propensity scoring [\cite{rosenbaumrubin1983}],
which rely
on auxiliary information available for the full sampling frame to estimate
unknown sampling probabilities. Link-tracing differs from
the traditional context of such methods in that the sampling
probabilities are
unobserved even when the design is executed faithfully, and in that the unknown
sampling probabilities result directly from the unobserved variable of interest.
In particular, estimating unknown sampling probabilities is equivalent to
estimating unobserved relations based on the observed relations. One
approach is to augment the sample with sufficient information to allow
for determination of the sampling probabilities. However in most cases,
this requires a substantial expansion of the sampling design.
Therefore, in practice we must rely on a model relating the observed
portions of the
network structure to the unobserved portions. Lack of reliance on an assumed
outcome model is a great advantage of the design-based framework over
the model-based 
framework. By introducing a model to estimate sampling probabilities
based on the outcome of interest,
we reintroduce this reliance on model form, negating much of the
advantage of the design-based framework. Furthermore, note that the naive
use of this approach has an ad-hoc flavor, while still requiring
complex observation weights and variance estimators.

In the next section, we describe an alternative more flexible
likelihood 
approach to network inference based on link-tracing samples.





\subsection{Likelihood-based inference}\label{modelinference}

Consider a parametric model for the random behavior of $\itall$
depending on a parameter $p$-vector $\veta$:
%
\begin{eqnarray}
P_{\veta}(Y=y),\qquad       \veta\in\Xi.
\end{eqnarray}
In the model-based framework, if $\itall$ is completely observed,
inference for $\veta$ can be based on the likelihood
\[
L[\veta\vert Y=y ]\propto P_{\veta}( Y=y).
\]
This situation has been considered in detail in \citet{HunHan06} and the
references therein. In the general case, where $Y$ may be only partially
observed, we can consider using
the (so-called) \textit{face-value likelihood} based solely on $Y_{\mathrm{obs}}$:
%
\begin{equation}
L[\veta\vert Y_{\mathrm{obs}} = y_{\mathrm{obs}} ]\propto\sum_{v \in{\cal
Y}(y_{\mathrm{obs}})} {P_\veta(Y=y_{\mathrm{obs}}+v)}.
\label{facevalue}
\end{equation}
%
This ignores the additional information about $ \veta  $ available in $D.$
Inference for $ \veta  $ and $ \psi $ should be based on all the available
observed data, including the sampling design information. This
likelihood is
any function of $\veta$ and $\psi$ proportional to
$P(D, Y_{\mathrm{obs}} \vert\veta,\psi)$:
\begin{eqnarray*}
  &&  L[\veta,\psi\vert Y_{\mathrm{obs}}=y_{\mathrm{obs}}, D=d_{\mathrm{obs}}]\\
&&\qquad\propto P(D=d_{\mathrm{obs}}, Y_{\mathrm{obs}}=y_{\mathrm{obs}} \vert\veta, \psi)\\
&&\qquad=\sum_{v \in{\cal Y}(y_{\mathrm{obs}})}
{P(D=d_{\mathrm{obs}} \vert Y=y_{\mathrm{obs}}+v, \psi)}{P_\veta(Y=y_{\mathrm{obs}}+v)}.
\end{eqnarray*}
Thus the correct model is related to the complete data model through the
sampling design as well as the observed nodes and dyads.

In likelihood inference, 
the sampling parameter $\psi$ is a nuisance parameter, and modeling
the sampling design along with the data structure adds a great deal of
complexity. It is natural to ask when we might consider the simpler
face-value likelihood, (\ref{facevalue}),
which ignores the sampling design.

In the context of missing data, \cite{Rub76} introduced the
concept of \textit{ignorability} to specify when inference based
on the face-value likelihood is efficient. We introduce the term
\textit{amenability} to represent the notion of ignorability for
network sampling strategies within a likelihood framework. 

In many situations where models are used, the parameters
$\veta\in\Xi$ and $\psi\in\Psi$ are \textit{distinct}, in the
sense that the joint parameter space of $ ({\veta
,\psi} )$ is $\Psi\times\Xi.$
If the sampling design is adaptive
and the parameters $ \veta $ and $ \psi $ are distinct,
\begin{eqnarray*}
  &&L[\veta,\psi\vert Y_{\mathrm{obs}} =y_{\mathrm{obs}}, D=d_{\mathrm{obs}}]\\
&&\qquad\propto P(D=d_{\mathrm{obs}} \vert Y_{\mathrm{obs}}
=y_{\mathrm{obs}}, \psi) \sum_{v \in{\cal Y}(y_{\mathrm{obs}})}
{P_\veta(Y=y_{\mathrm{obs}}+v)}\\
&&\qquad\propto L[\psi\vert D=d_{\mathrm{obs}},
Y_{\mathrm{obs}} =y_{\mathrm{obs}} ]{\times}L[\veta\vert Y_{\mathrm{obs}} =y_{\mathrm{obs}} ].
\end{eqnarray*}

Thus if the sampling design is adaptive and the structural and
sampling parameters are distinct, then the sampling design is
\textit{ignorable} in the sense that the resulting likelihoods
are proportional. When this condition is satisfied
likelihood-based inference for $\veta,$ as proposed here, is
unaffected by the (possibly unknown) sampling design.
This leads to the following definition and result.

\begin{Definition*}
Consider a sampling design governed by
parameter $\psi\in\Psi$ and a
stochastic network model $P_{\veta}(Y=y)$ governed by
parameter $\eta\in\Xi$. We call the sampling design \textit{amenable to
the model} if the sampling design is adaptive
and the parameters $\psi$ and $\eta$ are distinct.
\end{Definition*}

\begin{Result*} Consider networks produced
by the stochastic network model $P_{\veta}(Y=y)$ governed by
parameter $\eta\in\Xi$ which are observed by a sampling design
with parameter $\psi\in\Psi$
amenable to the model.
Then the likelihood for $\veta$ and $\psi$ is
%
\[
L[\veta,\psi\vert Y_{\mathrm{obs}}=y_{\mathrm{obs}}, D=d_{\mathrm{obs}}]
\propto L[\psi\vert D=d_{\mathrm{obs}} ,
Y_{\mathrm{obs}} =y_{\mathrm{obs}}]{\times}L[\veta\vert
Y_{\mathrm{obs}}=y_{\mathrm{obs}} ].
\]

Thus likelihood-based inference for $ \veta$ from $ L[\veta
,\psi\vert Y_{\mathrm{obs}}, D] $ will be the same as likelihood-based
inference for $ \veta$ based on $ L[\veta\vert Y_{\mathrm{obs}} ].$

This result shows for standard designs such as the ego-centric,
single wave and multi-wave sampling designs in Section~\ref{mbs},
likelihood-based inference can be based on the face-value
likelihood $ L[\veta\vert Y_{\mathrm{obs}} ].$
This was first noted in the foundational paper of
\cite{thompsonfrank2000}.
Explicitly, this is
%
\[
L[\veta\vert Y_{\mathrm{obs}} =y_{\mathrm{obs}} ] \propto P( Y_{\mathrm{obs}} =y_{\mathrm{obs}} \vert
\veta)
= \sum_{v \in{\cal Y}(y_{\mathrm{obs}})}{P_\veta(Y=y_{\mathrm{obs}} + v)}.
\]
%
Hence we can evaluate the likelihood by just enumerating the
full data likelihood over all possible values for the missing data.

We may also wish to make inference about the design parameter
$\psi.$
The likelihood for $\psi$ based on
the observed data is any function of $ \psi $ proportional to $ P(D, Y_{\mathrm{obs}}
\vert\psi).$ For designs amenable to the model this is
\begin{eqnarray*}
L[\psi\vert D=d_{\mathrm{obs}}, Y_{\mathrm{obs}}=y_{\mathrm{obs}} ]
&\propto& P(D=d_{\mathrm{obs}}\vert
Y_{\mathrm{obs}}=y_{\mathrm{obs}}, \psi)\\
&=&
P(D=d_{\mathrm{obs}}\vert Y=y_{\mathrm{obs}}+v, \psi)
\end{eqnarray*}
for any choice of $v$ in ${\cal Y}(y_{\mathrm{obs}}).$
Hence it can be computed without reference to the network model.
\end{Result*}

\section{Exponential family models for networks}\label{sec4}
The models we consider for the random behavior of $Y$ rely on a
$p$-vector $\vZ(Y)$ of statistics and a parameter vector
$\veta\in R^p$.
The canonical exponential family model is
%
\begin{eqnarray}
P_{\veta}(Y=y) = \exp\{\veta\cdot \vZ(y)-\kappa(\veta)\},\qquad
  y\in{\cal
Y}
\label{ergm}
\end{eqnarray}
where
$\exp\{\kappa(\veta)\} =
\sum_{u\in{\cal Y}}\exp\{\veta\cdot \vZ(u)\}
$
is the familiar normalizing constant associated with an
exponential family of distributions [\cite{Bar78}; \cite{Lehmann1983}].

The range of network statistics that might be included in the $\vZ(y)$
vector is vast---see \cite{wassermanfaust1994} for the most
comprehensive treatment of these statistics---though we will consider
only a few in this article. We allow
the vector $\vZ(y)$ to include covariate information about nodes or
edges in the graph in addition to information derived directly from
the matrix $y$ itself.

There has been a great deal of work on models of the form~(\ref{ergm}), to which we refer as exponential family random graph models
or ERGMs for short. %
[We avoid the lengthier EFRGM, for ``exponential
family random graph models,'' both for the sake of brevity and because
we consider some models in this article that should technically be called
\textit{curved} exponential families \citet{HunHan06}.]
%

The normalizing constant is usually difficult to compute directly for
${\cal Y}$ containing large numbers of networks.
Inference for this class of models was considered in
the seminal paper by \citet{Gey92}, building
on the methods of \cite{Fra86}
and the above cited papers.
Until recently, inference for social network models
has relied on\break maximum pseudolikelihood estimation \
[\cite{Bes74}; \ \cite{Fra86}; \cite{Str90}; \cite{Gey92}].\break
\citet{Gey92}
proposed a stochastic algorithm to approximate maximum likelihood
estimates for model~(\ref{ergm}), among other models; this
Markov chain Monte Carlo (MCMC) approach forms the basis
of the method described in this article.
The development of these methods for social network data has been
considered by
\citet{Cor98}; \cite{Cro98}; \cite{Sni02};\break
\cite{Han02}; \cite{Cor02};\break \cite{HunHan06}.

\subsection{Likelihood-based inference for ERGM}\label{sec:Inference}

In this section we consider likelihood inference for
$\veta$ in the case where $Y=Y_{\mathrm{obs}}+Y_{\mathrm{mis}}$
is possibly only partially observed.

As the direct computation of the likelihood is difficult when the
number of
networks in ${\cal Y}$ is large, we can approximate the likelihood by
using the MCMC approach of randomly sampling from the space of possible values
of the missing data and taking the mean. Alternatively, consider the
conditional distribution of $Y$ given $Y_{\mathrm{obs}}$:
\[
P_\veta(Y_{\mathrm{mis}} =v\vert Y_{\mathrm{obs}} =y_{\mathrm{obs}}) = \exp
 [{\veta\cdot \vZ(v+y_{\mathrm{obs}})-\kappa(\veta\vert y_{\mathrm{obs}}
)} ],\qquad    v\in{\cal Y}(y_{\mathrm{obs}}),
\label{eqn:cdist}
\]
where
%
$\exp [ \kappa(\veta\vert y_{\mathrm{obs}}) ]
=\sum_{u \in{\cal Y}(y_{\mathrm{obs}})} {\exp [
{\veta\cdot \vZ(u+y_{\mathrm{obs}})}  ]}.$
This formula gives a
simple way to sample from the conditional distribution and hence
produce multiple imputations of the full data. Specifically,
the conditional distribution of $Y$ given $Y_{\mathrm{obs}}$
is an ERGM on a constrained space of networks, and
hence one can simulate from it using a variant of the standard MCMC for ERGM
[\citet{HunHan06}; \cite{statnet}]
that restricts the proposed networks to the
subset of networks that are concordant to the observed data.

Also note that
\[
L[\veta\vert Y_{\mathrm{obs}} =y_{\mathrm{obs}} ]\propto
\exp [ \kappa(\veta\vert y_{\mathrm{obs}}) - \kappa(\veta) ]
\]
which can then be estimated by MCMC samples: the first term by a chain
on the
complete data and the second by a chain conditional on $ y_{\mathrm{obs}} $. So
the sampled data situation is only slightly more difficult than the complete
data case.


\section{Two-wave link-tracing samples from a collaboration network}\label{sec5}


In this section we investigate the effect of network sampling on estimation
by comparing network samples to the situation where we observe the complete
network.
Specifically, we consider the collaborative
working relations between 36 partners in a New England law
firm introduced in Section~\ref{intro}. These data have been studied by many authors
including
\cite{laz01}, \cite{sprh04} and \cite{HunHan06} (whom we follow).

\begin{table}[b]
\tablewidth=295pt
\caption{Bias and Root Mean Squared Error $($RMSE$)$ of natural
parameter MLE based on two-wave samples as percentages of true
parameter values and efficiency losses}\label{tab:Table1}
\begin{tabular*}{295pt}{@{\extracolsep{\fill}}lcccc@{}}
\hline
\textbf{Natural} & \textbf{Complete} &  \textbf{Bias}  &  \textbf{RMSE}
&  \textbf{Efficiency}\\
\textbf{parameter} &\textbf{data  value} & \textbf{(\%)} &\textbf{(\%)} &\textbf{loss (\%)} \\
\hline
\textit{Structural}\\
 Edges & $-$6.51 & 0.2 & 1.2& 1.7 \\
 GWESP &   \phantom{$-$}0.90 & 0.8& 3.7& 5.1 \\[3pt]
\textit{Nodal}\\
 Seniority &  \phantom{$-$}0.85 & 0.3 & 3.1& 1.3 \\
 Practice &   \phantom{$-$}0.41 & 0.4 & 5.3& 3.5 \\[3pt]
\textit{Homophily}\\
 Practice &   \phantom{$-$}0.76 & 0.8& 4.3& 2.9 \\
 Gender &   \phantom{$-$}0.70 & 0.9& 4.7& 1.7 \\
 Office &   \phantom{$-$}1.15 & 0.7& 2.9& 2.8 \\
\hline
\end{tabular*}
\end{table}

We consider an ERGM (\ref{ergm}) with two network statistics
for the direct effects of seniority and practice of the form
\[
\sum_{1\le i,j\le n}y_{ij}X_i ,
\]
where $X_i$ is the seniority or practice of partner $i.$
We also consider three dyadic homophily attributes based on
practice, gender and office.
These are included as three
network statistics indicating matches between the two partners in the
dyad on
the given attribute:
\[
\sum_{1\le i<j\le n}y_{ij}{\cal I}(X_i = X_j),
\]
where ${\cal I}(x)$ indicates the truth of the condition $x$ and $X_i$ and
$X_j$ are the practice, gender or office attribute of partner $i$ and
$j,$ respectively.
We also include statistics that are purely functions of the relations $y$.
These are the number of edges
(essentially the density) and the geometrically weighted edgewise
shared partner statistic (denoted by GWESP), a measure of the
transitivity structure in the network [\cite{sprh04}].
The model is a slightly reparameterized form of Model~2
in \cite{HunHan06} obtained
by replacing the alternating $k$-triangle
term with the GWESP term.
The scale parameter for the GWESP term is fixed
at its optimal value (0.7781).
See \cite{HunHan06} for details.

As discussed in \cite{HunHan06}, this model provides an adequate fit
to the
data, and we will use it here to assess the effect of sampling
on model fit.
A~summary of the MLE parameters used is given in the \textit{complete data
value} column of Table~\ref{tab:Table1}. Note that we are taking these parameters as ``truth'' and
considering data produced by sampling from this network.

\begin{figure}

\includegraphics{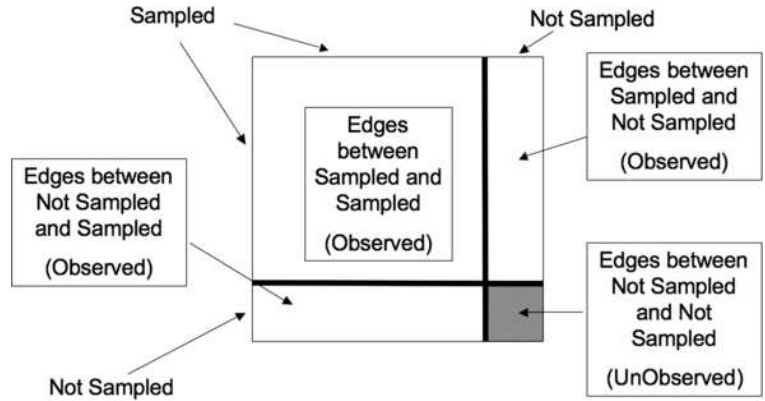}

\caption{Schematic depiction of sampled and unobserved arc data when the
sampling is over an undirected network.} \label{fig:sampblock2by2}
\end{figure}

We construct all possible datasets produced
by a two-wave link-tracing design starting from two randomly
chosen nodes (the ``seeds'').
This adaptive design is amenable to the model.
As there are 36 partners and the sample is
deterministic given the seeds, there
are ${36\choose2} = 630$ possible data sets.
The number of actors in each dataset varies from just $2$ to all $36$
depending on the degree of connectedness of the seeds.
The data pattern is shown in Figure~\ref{fig:sampblock2by2}.
Consider a partition of the sampled from the nonsampled and the
corresponding $2 \times2$ blocking of the sociomatrix, with the
four blocks representing dyads from sampled and
nonsampled to sampled and nonsampled.
The complete data consists of the full sociomatrix.
The first three blocks contain the observed data, the dyads involving
at least one sampled node,
and the last block contains the unobserved data, those
between the nonsampled.

For each of these samples we use the methods of Section~\ref
{sec:Inference} to
estimate the parameters. We can then compare them to the MLE for the complete
dataset.
For these networks, the MLEs
are obtained using \texttt{statnet}
[\citet{statnet}], both for the natural parametrization and for the
mean value parameterization 
[see \citet{Han03}].

The mean value
parameters are a function of the natural parameters, specifically the expected
values of the sufficient statistics given the values of the natural parameters.

There are two isolates, that is nodes with no relations. If these two
are selected as the two seeds, only 69 of the 630 dyads are observed,
and no edges are observed. Therefore, the MLE associated with this
sample includes (negative) infinite values, on the boundary of the
convex hull.
For this reason, we exclude this sample from our analyses.
Practically, this exclusion is reasonable in that it is unlikely any
researcher drawing a link-tracing sample including only two isolated
nodes will proceed with analysis of that sample.


One way to assess the effect of the link-tracing design is to compare
the estimates from
the sampled data to that of the complete data. As a measure of the
difference between the estimates in the metric of the model, we use the
Kullback--Leibler divergence
from the model implied by the complete data estimate to that of the
sampled data estimate. Recall that the Kullback--Leibler divergence of a
distribution with probability mass
function $p$ from the distribution with probability mass function $q$
is
\[
E_q[\log(q)-\log(p)].
\]
Let $\veta$ and $\vxi$ be alternative
parameters for the model (\ref{ergm}).
The Kullback--Leibler divergence, $\operatorname{KL}(\vxi,\veta)$,
of the ERGM with parameter $\veta$ from
the ERGM with parameter $\vxi$ is
\begin{eqnarray*}
E_{\xi} \biggl[\log \biggl(\frac{P_{\xi}(Y=y)}{P_{\veta}(Y=y)}
\biggr)  \biggr] &=&
\sum_{y \in\mathcal{Y} } \log \biggl(\frac{P_{\xi}(Y=y)}{P_{\veta
}(Y=y)} \biggr)
P_{\xi}(Y=y)\\
&=& \sum_{y \in\mathcal{Y} } (\vxi- \veta)\cdot yP_{\vxi}(Y=y) +
\kappa(\veta) - \kappa(\vxi)\\
&=& (\vxi- \veta)\cdot E_{\xi}[\vZ(Y)] +
\kappa(\veta) - \kappa(\vxi).
\end{eqnarray*}
If $\vxi$ is the complete data MLE then $E_{\xi}[\vZ(Y)] = \vZ
(Y_{\mathrm{obs}})$ are the observed statistics (given in the \textit{complete data
value} column of Table~\ref{tab:Table2}).
The divergence can be easily computed using the MCMC algorithms of Section~\ref{sec:Inference}.

\begin{table}[b]
\caption{Bias and Root Mean Squared Error $($RMSE$)$ of mean value
parameter MLE based on two-wave samples as percentages of true
parameter values and efficiencies}\label{tab:Table2}
\begin{tabular}{@{}lcccc@{}}
\hline
\textbf{Natural} &\textbf{Complete} & \textbf{Bias} & \textbf{RMSE}
&\textbf{Efficiency}\\
\textbf{parameter} &\textbf{data  value} & \textbf{(\%)} &\textbf{(\%)} &\textbf{loss (\%)} \\
\hline
 \textit{Structural}\\
 Edges & 115.00 & 0.4 & 2.0 & 1.8 \\
 GWESP & 190.31 &0.4 & 2.8 & 1.9 \\[3pt]
 \textit{Nodal}\\
 Seniority &130.19 & 0.3 & 1.8 & 1.4 \\
 Practice & 129.00 & 0.2 & 2.6 & 3.4 \\[3pt]
\textit{Homophily}\\
 Practice & \phantom{1}72.00 & 0.1 & 2.0 & 1.7 \\
 Gender & \phantom{1}99.00 & 0.5 & 2.1 & 1.8 \\
 Office & \phantom{1}85.00 & 0.7 & 2.7 & 3.0 \\
\hline
\end{tabular}
\end{table}

\begin{figure}

\includegraphics{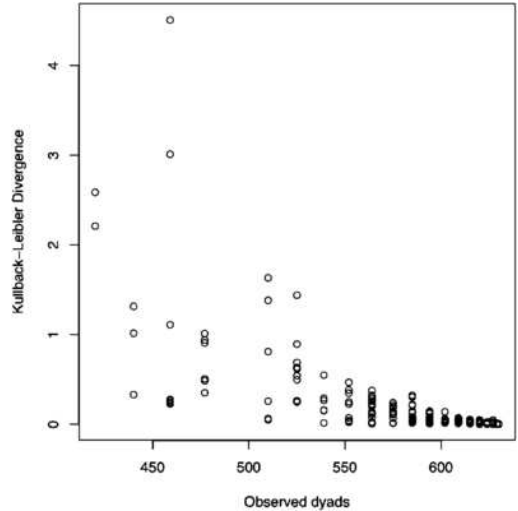}

\caption{Kullback--Leibler divergence of the MLEs based on the
samples compared to the complete data MLE. As the number of dyads
sampled increases, the information content of the samples
approaches that of the complete data. The information loss for
the majority of samples is modest.}
\label{fig:ndyads}
\end{figure}

Figure~\ref{fig:ndyads} plots the Kullback--Leibler divergence of the
MLEs based on the
629~samples from the complete data MLE. The Kullback--Leibler
divergence of the two smallest 
samples,
including only 5 nodes (165 dyads), are about 14 and have not been
plotted to reduce the vertical scale.
The horizontal axis is the number of observed dyads in the sample.
The plot indicates how the
information in the data about the complete data MLE approaches
that of the complete data as the number of sampled dyads approaches the
full number.
The key feature of this figure is the \textit{variation} in information content
among samples of the same size especially for the smaller sample sizes.
Different seeds lead to samples that tell us different things about the
model even
when the numbers of partners surveyed is the same.

For more specific information on the individual estimates,
we can compute the
bias of the estimates based on the samples as the mean difference
between the
parameter estimates from the samples and that of the complete network.
The root mean squared error (RMSE) is the square-root of the mean
of the squared difference between the parameter
estimates from each sample and the complete data estimates.
The efficiency loss of the sampled estimate is the ratio of the
mean squared error and the variance of the sampling distribution of the
estimate based on
the full data. This standardizes the error in the sampled estimates by
the variation in the complete data estimates.
We also complete a similar comparison of the
estimates under the alternative mean value parametrization
[\citet{Han03}].\looseness=1

The properties of the 
natural parameter estimates
are summarized in Table \ref{tab:Table1}. The bias and
root mean squared error are presented in percentages of the complete data
parameter estimates.

The bias is very small and the RMSE
is modest. The efficiency loss is 2\%--3\% on average. Note that these
population-average
figures obscure the variation in loss over individual samples apparent
in Figure~\ref{fig:ndyads}.


Table \ref{tab:Table2} is the mean value parameterization analog of
Table~\ref{tab:Table1}. As these are on
the same measurement scale as the statistics they are easier to interpret.
Again we see that the estimates are approximately unbiased and the RMSE and
efficiency losses are small. 


\section{Discussion}\label{sec6}

In this paper we give a concise and systematic statistical framework
for dealing with
partially observed network data resulting from a designed sample. The
framework includes, but is not restricted to,
adaptive network sampling designs.
We present a definition of a network design which is amenable to a
given model
and a result on likelihood-based
inference under such designs.

An important simple result of this framework is that
sampled networks are not ``biased'' but can be representative if analyzed
correctly. Many authors have confused the ideas of simple random
sampling of the dyads
with representative designs. The results of this paper indicate that
simple random sampling is not
necessary for valid inference. In fact, the most commonly used
designs can be easily taken into account.
Hence, despite their form, inference from adaptive network samples
is tractable.

It is illustrative to compare our approach to that of \cite{stumpfwm05}.
These authors highlight the difference between the structure of a
network and that of a sub-network induced by Bernoulli
sampling of its nodes.
The framework in this paper allows valid inference for the
properties of the network based on its partial observation.
This is because we fit a broad class of models compatible with
an arbitrary set of network statistics (e.g., ERGM) for the complete network
and use a method of inference that does not rely on equality between
the structure of the full and sub-networks.
As illustrated by the work of \cite{stumpfwm05}, treating the
observed portion as if it were the full network may lead to invalid
inference about characteristics of the full network such as
the degree distribution.


We have also shown that likelihood-based inference from an adaptive
network sample can be conducted
using a complete network model. We
have shown that such inference is both principled and practical.
The likelihood framework naturally accommodates standard
sampling designs.
Note that in a design-based frame, principled inference
would require a great deal of effort to precisely characterize
the sampling designs.
%
The result that link-tracing designs are adaptive and can be analyzed
with complex likelihood based methods is very valuable in practice as these
designs have previously not been analyzed with general exponential family
random graph (or similar) models. The only prior work appears to be
that of \cite{thompsonfrank2000} who applied a less complex model class.


In our application we show that an adaptive network sampling of a
collaboration network can lead to effective estimates of the
model parameters in the vast majority of cases. We find that the
MLEs from the samples have only modest bias (compared to the
complete data estimate) and an error that only increases slowly
with the number of unobserved dyads. We also show that the
information content of the sample (with respect to the model),
varies greatly even for samples of the same size. For
conventional samples of i.i.d. random variables, the Fisher
information is simply proportional to the sample size. In the
network setting with dependence terms, however, the Fisher
information will depend on the specific set of nodes and dyads sampled.
For example, the information component corresponding to the GWESP term
in the example
will be larger for samples in which more pairs of nodes joined by edges
are sampled, as GWESP applies only to pairs
of nodes joined by edges.
If no such dyads were sampled, there would be no information in the
sample about the propensity for nodes joined
by edges to have relations in common.



In practice the sample is a result of a combination of
the sampling design and an
\textit{out-of-design }mechanism.
The sampling design is the part of the observation process
under the control of the surveyor.
When adaptive designs are executed faithfully, the unknown dyads are
assumed to be intentionally unobserved, or
missing by design. Note that the definition of control may be extended
to nonamenable sampling designs, for example
by allowing the design to depend on unknown factors,
such as the unrecorded values of variables used for stratification.
The out-of-design mechanism is the
nonintentional nonobservation of network information
(e.g., due to the failure to report links, incomplete measurement of
links and attrition from longitudinal surveys).
This is also referred to, in general, as the
\textit{non-response mechanism.}
We consider the joint effect of sampling and missing data in a
companion paper
[\citet{HanGile07}].




%

\section*{Acknowledgments}
The authors would like to thank the members of the UW Network Modeling
Group (Martina Morris, P. I.),
Stephen Fienberg and the reviewers for their helpful input.

\begin{supplement}[id=suppA]
\sname{Supplement}
\stitle{Software used in the simulation study\\}
\slink[doi]{10.1214/08-AOAS221SUPP}
\slink[url]{http://lib.stat.cmu.edu/aoas/221/supplement.zip}
\sdatatype{.zip}
\sdescription{The code used to perform this study is written in the
\texttt{R}
statistical language [\textsf{R} Development Core Team (\citeyear{R})] and is
based on
\texttt{statnet}, an open-source
software suite for network modeling [Handcock et al. (\citeyear{statnet})].
We provide the code and documentation for it with links to the
\texttt{statnet} website.}
\end{supplement}

\printaddresses

\end{document}